\documentclass[a4paper,fleqn,usenatbib]{mnras}

\usepackage[T1]{fontenc}
\usepackage{ae,aecompl}

\usepackage{graphicx}	
\usepackage{amsmath}	
\usepackage{amssymb}	

\usepackage[usenames,dvipsnames]{color}

\DeclareMathAlphabet{\mathpzc}{OT1}{pzc}{m}{it}
 
\def\beq{\begin{equation}}
\def\eeq{\end{equation}}
\def\bea{\begin{eqnarray}}
\def\eea{\end{eqnarray}}
\def\bwt{\begin{widetext}}
\def\ewt{\end{widetext}}
\def\kms{~{\rm km\cdot s^{-1}}}
\def\mpc{{\rm Mpc^{-1}}}
 
\def\nn{\nonumber\\}

\title[Imprint of $f(R)$ gravity in the cosmic magnification]{Imprint of $f(R)$ gravity in the cosmic magnification}

\author[Duniya, Abebe, de la Cruz-Dombriz, Dunsby]{
Didam G. A. Duniya,$^{1}$\thanks{E-mail: duniyaa@biust.ac.bw}
Amare Abebe,$^{2,3}$
\'Alvaro de la Cruz-Dombriz$^{4,5}$ and
\newauthor\ Peter K. S. Dunsby$^{5}$
\\
$^{1}$Department of Physics \& Astronomy, Botswana International University of Science and Technology, Palapye, Botswana\\
$^{2}$Centre for Space Research, North-West University, South Africa\\
$^{3}$National Institute for Theoretical and Computational Sciences (NITheCS), South Africa\\
$^{4}$Departamento de F\'isica Fundamental, Universidad de Salamanca, P. de la Merced, 37008 Salamanca, Spain\\
$^{5}$Department of Mathematics \& Applied Mathematics, University of Cape Town, Cape Town 7701, South Africa
}

\date{Accepted XXX. Received YYY; in original form ZZZ}

\pubyear{2022}

\begin{document}
\label{firstpage}
\pagerange{\pageref{firstpage}--\pageref{lastpage}}
\maketitle

\begin{abstract}
$f(R)$ gravity is one of the simplest viable modifications to General Relativity: it passes local astrophysical tests, predicts both the early-time cosmic inflation and the late-time cosmic acceleration, and also describes dark matter. In this paper, we probe cosmic magnification on large scales in $f(R)$ gravity, using the well-known Hu-Sawicki model as an example. Our results indicate that at redshifts $z \,{<}\, 3$, values of the model exponent $n \,{>}\, 1$ lead to inconsistent behaviour in the evolution of the scalar perturbations. Moreover, when relativistic effects are taken into account in the large scale analysis, our results show that as $z$ increases, large-scale changes in the cosmic magnification angular power spectrum owing to integral values of $n$ tend to share a similar pattern, while those of decimal values tend to share another. This feature could be searched for in the experimental data, as a potential ``smoking gun" for the given class of gravity models. Furthermore, we found that at $z \,{=}\, 1$ and lower, relativistic effects lead to a suppression of the cosmic magnification on large scales in $f(R)$ gravity, relative to the concordance model; whereas, at $z \,{>}\, 1$, relativistic effects lead to a relative boost of the cosmic magnification. In general, relativistic effects enhance the potential of the cosmic magnification as a cosmological probe.
\end{abstract}

\begin{keywords}
Cosmology -- (cosmology:) dark energy < Cosmology -- cosmology: theory < Cosmology
\end{keywords}



\section{Introduction}\label{sec:intro}
A key problem in modern cosmology is to identify the cause of the accelerated expansion of the Universe at late cosmic times. The lack of a fundamental understanding within General Relativity for this accelerated expansion has led to alternative theories of gravity, referred to as modified gravity (MG) \citep[see e.g.][]{Hu:2007nk, Starobinsky:2007hu, Tsujikawa:2007xu, Frolov:2008uf, Cembranos:2008gj, Amendola:2010bk, Nojiri:2008nt, Clifton:2011jh, Clifton:2015ira, Katsuragawa:2016yir, Katsuragawa:2017wge, Ishak:2018his, Chen:2019kcu, MacDevette:2022hts, Duniya:2019mpr}. 
\citep[See particularly,][for extensive reviews on MG.]{Clifton:2011jh,  Ishak:2018his}

One of the most widely studied theories is one where the Lagrangian density is given as a function of the Ricci scalar $R$, which is commonly known as the $f(R)$ theory \citep[see e.g.][]{Hu:2007nk, Starobinsky:2007hu, Tsujikawa:2007xu, Frolov:2008uf, Cembranos:2008gj, Amendola:2010bk, Nojiri:2008nt, Clifton:2011jh, Clifton:2015ira, Katsuragawa:2016yir, Katsuragawa:2017wge, Ishak:2018his, Chen:2019kcu, MacDevette:2022hts}. This theory is one of the simplest modifications to General Relativity; it passes local astrophysical tests, predicts both the early-time cosmic inflation and the late-time cosmic acceleration, and also describes dark matter \citep[see e.g.][]{Cembranos:2008gj, Nojiri:2008nt, Katsuragawa:2016yir, Katsuragawa:2017wge, Chen:2019kcu}.

While (traditional) dark energy models \citep[see e.g.][]{Amendola:2010bk, Duniya:2013eta, Duniya:2015nva, Duniya:2015ths, Duniya:2015dpa, Duniya:2016gcf} are intended to describe the late-time cosmic accelerated expansion, it is important for any viable MG model to describe both the late-time cosmic acceleration and the strong-field gravity limit in the solar system and other contexts, such as gravitational-wave emission and neutron-star phenomenology \citep[see e.g.][]{Barack:2018yly}. Until recently, it was unclear in the literature whether the class of suggested $f(R)$ models in the metric formalism were able to satisfy the strong solar-system conditions and still cause the late-time acceleration of the cosmic expansion, without a cosmological constant $\Lambda$. Some of the conditions under which $f(R)$ models in the metric formalism are cosmologically viable include \citep[see e.g.][]{Amendola:2010bk, MacDevette:2022hts}: (1) $1+\partial f(R)/\partial{R} \,{>}\, 0$ for $R \,{\geq} R_0$ (with $R_0$ being the value of the Ricci scalar at the present epoch), which is required to avoid anti-gravity behaviour; (2) $\partial^2 f(R)/\partial{R}^2 \,{>}\, 0$ for $R \,{\geq} R_0$, which is required for both the consistency with solar-system gravity constraints and the presence of a matter-dominated epoch; and (3) $f(R) \,{\to}\, {-}2\Lambda$ for $R \,{\gg}\, R_0$, i.e. at very high redshifts. 

In recent years \cite{Hu:2007nk}, \cite{Starobinsky:2007hu} and \cite{Tsujikawa:2007xu} (amongst other authors), proposed $f(R)$ models in the metric formalism that satisfy all the aforementioned conditions; yet leading to the accelerated cosmic expansion at late times and, satisfying solar-system gravity requirements (in the weak-field limit). 

Although the $f(R)$ gravity has been extensively studied, yet its effects on the cosmic magnification \citep[e.g.][]{Blain:2001yf, Schneider:2006eta, LoVerde:2006cj, Ziour:2008awn, Bonvin:2008ni, Schmidt:2009rh, Schmidt:2010ex, Jeong:2011as, Raccanelli_2012, Raccanelli:2013gja, Raccanelli:2016avd, Liu:2013yna, Camera:2013fva, Bacon:2014uja, Duniya:2015ths, Duniya:2016ibg, Duniya:2016gcf, Hildebrandt:2015kcb, Montanari:2015rga, Bonvin:2016dze, Chen:2018hil, Andrianomena:2018aad, Ballardini:2018cho, Baklanov:2020edg, Liao:2020yqz, Chan:2021jhh, Bayer:2021ugw, Baldwin:2021lud, Er:2022lad}, which is an important phenomenon in cosmology, has never been investigated for this class of MG. For instance, cosmic magnification will be crucial in interpreting the data from forthcoming HI surveys of the SKA \citep{Blake:2004pb, Maartens:2015mra} and the baryon acoustic oscillation surveys of BOSS \citep{SDSS:2011jap, BOSS:2012dmf}. Also, the cosmic magnification will be key to understanding both cosmic distances and the geometry of the Universe. Moreover, forthcoming surveys in the optical and the radio bands will extend to large cosmic scales, at the survey redshifts; on these scales, relativistic effects \citep[see e.g.][]{Bonvin:2011bg, Jeong:2011as, Duniya:2015ths, Montanari:2015rga, Duniya:2015ths, Duniya:2016ibg, Duniya:2015dpa, Duniya:2016gcf, Raccanelli:2013gja, Duniya:2019mpr, Duniya:2022xcz} become significant. With the expected precision of forthcoming experiments, surveys on these scales will provide the best constraints on alternative theories of gravity, and probe the imprint of relativistic effects. Thus, theoretical work needs to be done in view of identifying the imprint of these theories, in the cosmological observables.

Using the well-known Hu-Sawicki model \citep{Hu:2007nk} as a case study, we investigate cosmic magnification in $f(R)$ gravity, using the magnification angular power spectrum, on large scales; taking full account of relativistic effects in the observed magnification density contrast. Such a study will help us to understand whether relativistic effects may be important in identifying signatures of $f(R)$ gravity. This work mainly seeks to set a background basis for future tests of $f(R)$ gravity with the cosmic magnification, by performing a qualitative analysis of the imprint of $f(R)$ gravity in the magnification angular power spectrum. We start by outlining the relevant dynamics of the $f(R)$ gravity in Sec.~\ref{sec:fR}. In Sec.~\ref{sec:DeltaMag} we give the relativistic form of the observed cosmic magnification overdensity, while in Sec.~\ref{sec:MagCls} we discuss the magnification angular power spectrum in $f(R)$ gravity. We conclude in Sec.~\ref{sec:Conc}.


\section{The Universe with $\lowercase{f}(R)$ Gravity}\label{sec:fR}
In this section we consider the $f(R)$ theory of gravity, which admits time derivatives higher than second order. These higher derivatives are able to render the $f(R)$ gravity to be less susceptible to ghost-like instabilities, e.g. if the higher derivatives only act on modes that would otherwise remain non-dynamical---such as the conformal mode (which does not propagate) in general relativity---then these derivatives may only cause them to merely propagate, rather than rendering them as ghosts \citep{Clifton:2011jh, Ishak:2018his}. 

\subsection{Notation}
The equations we use in this paper are drawn from the rigorous work by \citet{Clifton:2011jh}. However, here we reinstate standard notations; with the Newton's gravitational constant $ G \,{\neq}\, 1$, the speed of light $c \,{=}\, 1$, and
\bea\label{ftrans}
f \to f + R, \quad F \to 1 + f_R,
\eea
where, as mentioned above, $R$ is the Ricci scalar, $f \,{=}\, f(R)$ is an arbitrary function (to be specified) of $R$, and $f_R \,{\equiv}\, \partial{f}/\partial\bar{R}$. Henceforth, $\bar{X}$ and $\delta{X}$ denote the background and the perturbation terms, respectively, for a given parameter $X$; with $|\delta{X}| \,{\ll}\, 1$.

We adopt a general flat spacetime metric, with the $(-,\, +,\, +,\, +)$ signature, given by
\bea\label{metric1}
ds^2 &=& g_{\mu\nu} dx^\mu dx^\nu, \nn
&=& a^2\left\{-(1+2\phi) d\eta^2 + 2B_i dx^i d\eta \right.\nn
&& \left.\hspace{0.6cm} +\; \left[(1-2D)\delta_{ij} + 2E_{ij}\right] dx^i dx^j \right\},
\eea
where we have parametrized the metric tensor perturbation $\delta{g}_{\mu\nu}$ by scalar-field degrees of freedom: $\phi \,{=}\, \phi(\eta,x^i)$, $B \,{=}\, B(\eta,x^i)$, $D \,{=}\, D(\eta,x^i)$, and $E \,{=}\, E(\eta,x^i)$; with $a \,{=}\, a(\eta)$ being the cosmic scale factor, $\eta$ being the conformal time, $x^i$ being the physical spatial coordinates, $B_i \,{=}\, \nabla_i B$ and $E_{ij} \,{=}\, \left(\nabla_i\nabla_j  - \frac{1}{3}\delta_{ij}\nabla^2\right) E \equiv D_{ij} E$; and $D_{ij}$ is a (spatial) longitudinal operator. However, given coordinate freedom (the condition that there are no preferred coordinates, with all physical laws retaining the same form in all coordinate systems) one is free to choose any coordinates. But by changing coordinates, the scalar perturbations also change.

In order to deal with this coordinate freedom, the coordinates are fixed by choosing a gauge. By taking a gauge transformation, which modifies the coordinate-dependent perturbations (without affecting the coordinates), we are able to define new potentials in (gauge-invariant) forms that preserve the physical equations, given by
\bea\label{DefnPhiPsi}
\Phi &\equiv & \phi - {\cal H}\left(E'-B\right) + B' - E'', \nn
\Psi &\equiv & D + \dfrac{1}{3}\nabla^2 E + {\cal H}\left(E'-B\right),
\eea
which are the well known as the Bardeen potentials \citep{Bardeen:1980kt, Bonvin:2011bg, Duniya:2015ths, Duniya:2016ibg, Duniya:2016gcf}; with ${\cal H} \,{=}\, a'/a$ being the comoving Hubble parameter, a prime denoting derivative with respect to $\eta$, and the 4-velocity corresponding to $\Psi$ and $\Phi$ being given by 
\bea\label{DefnV}
u^\mu = a^{-1}\left(1-\Phi, \nabla^i V\right),\quad\quad V = v+E',
\eea
where $v$ is the coordinate velocity potential.

We use the energy-momentum tensor for standard cosmic fluids, given by
\bea\label{T_munu}
T^0\/_0 &=& -\left(\bar{\rho} + \delta\rho_A\right),\nn
T^0\/_j &=& \left(\bar{\rho} + \bar{p}\right)\nabla_j V, \nn
T^i\/_j &=& \left(\bar{p}+\delta{p}\right)\delta^i\/_j + \left(\bar{\rho} + \bar{p}\right) D^i\/_j \Pi,
\eea
where $\bar{\rho}$ and $\bar{p}$ are the background energy density and pressure, respectively, and $\Pi$ is the anisotropic stress potential; with $V$ being the (gauge-invariant) velocity potential, as given by \eqref{DefnV}.


\subsection{The background equations}\label{subsec:backgrnd}
Here, we reformulate the $f(R)$ gravity in the dark energy (DE) paradigm \citep{Amendola:2010bk, Duniya:2013eta, Duniya:2015nva, Duniya:2015ths, Duniya:2015dpa, Duniya:2016gcf, Duniya:2019mpr}. We assume a late-time universe dominated by only two cosmic species: standard matter ($m$) and an effective DE ($x$), which is generated solely by the $f(R)$ gravity. 

It follows that the Friedmann equation is given by
\bea\label{Friedmann}
{\cal H}^2 = \dfrac{8\pi{G}a^2}{3}\left(\bar{\rho}_m + \bar{\rho}_x\right) \;\equiv\; \dfrac{8\pi{G}a^2}{3}\bar{\rho}_{\rm eff},
\eea
where ${\cal H}$ is as given in \eqref{DefnPhiPsi}; $\bar{\rho}_m$ is the background matter (energy) density, and the background energy density of the effective DE is given by
\bea\label{rho_DE}
8\pi{G} \bar{\rho}_x \equiv -3a^{-2}\left({\cal H}f'_R - {\cal H}'f_R\right) - \dfrac{1}{2}f,
\eea
where $f$ and $f_R$ are as given in \eqref{ftrans}. The associated acceleration equation, is given by 
\bea\label{acceleration}
{\cal H}' = -\dfrac{4\pi{G}a^2}{3}\left(\bar{\rho}_{\rm eff} + 3\bar{p}_{\rm eff}\right),
\eea
where $\bar{p}_{\rm eff} \,{=}\, \bar{p}_m \,{+}\, \bar{p}_x$ is the effective pressure of the system, and $\bar{p}_m$ is the background matter pressure; with the background DE pressure defined by
\bea\label{p_DE}
8\pi{G} \bar{p}_x \equiv a^{-2}\left(f''_R + {\cal H}f'_R\right) - a^{-2}\left({\cal H}' + 2{\cal H}^2\right)f_R + \dfrac{1}{2}f .
\eea
The matter and the DE background (density) evolution equations, are given by
\bea\label{rhoPrimes}
\bar{\rho}'_m + 3{\cal H}(1 + w_m)\bar{\rho}_m = 0,\quad \bar{\rho}'_x + 3{\cal H}\left(1 + w_x\right)\bar{\rho}_x = 0,
\eea
respectively, where $w_m \,{=}\, \bar{p}_m/\bar{\rho}_m$ is the matter equation of state parameter. (It is easy to show that the evolution of $\bar{\rho}_x$ as given by \eqref{rhoPrimes}, does hold.) For convenience, we introduce the following dimensionless (background) variables
\begin{align}\label{x_i}
&& x_1 \;{\equiv}\;& -\dfrac{f'_R}{{\cal H}f_R},& x_2 \;{\equiv}\;& -\dfrac{a^2f}{6{\cal H}^2f_R},\nn
&& x_3 \;{\equiv}\;& \dfrac{a^2\bar{R}}{6{\cal H}^2},& x_4 \;{\equiv}\;& -\dfrac{f_R}{1+f_R} ,
\end{align}
which lead to the background energy density parameters:
\bea\label{Omega_A}
\Omega_x &\equiv\;& \dfrac{8\pi{G}a^2}{3{\cal H}^2}\bar{\rho}_x = \left(1-x_1-x_2-x_3\right)\dfrac{x_4}{1+x_4},\nn 
\Omega_m &\equiv\;& \dfrac{8\pi{G}a^2}{3{\cal H}^2}\bar{\rho}_m = 1 - \Omega_x,
\eea
where we used \eqref{Friedmann} and \eqref{rho_DE}. The parameters in \eqref{x_i} evolve, respectively, according to  
\bea\label{x1_prime}
x'_1 &=& \left(x_1-x_3-3w_m\right) {\cal H}x_1 - 3(1+w_m)\left(x_2+x_3\right){\cal H} \nn
&& +\; \left(1-2x_3-3w_m\right)\dfrac{{\cal H}}{x_4}, \\\label{x2_prime}
x'_2 &=& \left(4+x_1-2x_3\right){\cal H}x_2 + x_1 \Gamma {\cal H}x_3, \\ \label{x3_prime}
x'_3 &=& 2\left(2-x_3\right){\cal H}x_3 - x_1 \Gamma {\cal H}x_3,\\ \label{x4_prime}
x'_4 &=& -x_1\left(1+x_4\right){\cal H}x_4,
\eea
with the Ricci scalar $\bar{R} = 6a^{-2}\left({\cal H}'+{\cal H}^2\right)$ and $\Gamma^{-1} \equiv d\log(f_R)/d\log(\bar{R}) = \bar{R} f_{RR}/f_R$. Thus, the equation of state parameter for the effective DE, $w_x \,{=}\, \bar{p}_x/\bar{\rho}_x$, is given by
\bea\label{w_de}
w_x \;=\; w_m + \dfrac{(2x_3 + 3w_m - 1)(1+x_4)}{3\left(x_3 + x_2 + x_1 - 1\right)x_4} , 
\eea
with $\bar{\rho}_x$ and $\bar{p}_x$ being given by \eqref{rho_DE} and \eqref{p_DE}, respectively.


\subsection{The perturbed field equations}\label{subsec:fieldEqs}
Here we reformulate the perturbations equations to correspond to the DE scenario in Sec.~\ref{subsec:backgrnd}. Thus, in a multi-component universe---with several cosmic species $A$---the gravitational field constraint equations, are given by
\begin{align} \label{dPsidt}
&& \Psi' + {\cal H}\Phi =& -4\pi{G}a^2\left(\bar{\rho}_{\rm eff} + \bar{p}_{\rm eff}\right) V_{\rm eff},\\ \label{grad2Psi}
&& \nabla^{2} \Psi - 3\mathcal{H} \left( {\cal H}\Phi +{\Psi}'\right) =\;& 4\pi{G}a^2 \delta\rho_{\rm eff},
\end{align}
where $V_{\rm eff}$ is the effective velocity potential of the system, given by
\bea\label{V_eff}\nonumber
V_{\rm eff} \;=\; \dfrac{1}{1+w_{\rm eff}} \sum_A \Omega_A\left(1+w_A\right)V_A,
\eea
with the DE velocity potential $V_x$, being given by (see Appendix~\ref{App:f(R)} for details)
\begin{align}\label{V_x}
3{\cal H}^2\Omega_x(1+w_x) V_x =\;& {\cal H}\delta{f}_R - 2\left(\Psi' + {\cal H}\Phi\right) \dfrac{x_4}{1+x_4} \nn
& +\; \dfrac{x_1 x_4}{1+x_4} {\cal H}\Phi - \delta{f}'_R, 
\end{align}
where $\Omega_x$ and $w_x$ are as given by \eqref{Omega_A} and \eqref{w_de}, respectively; $\delta\rho_{\rm eff} = \sum_A \delta\rho_A$ is the effective density perturbation of the system, and $\delta\rho_A$ is the coordinate density perturbation in $A$, with the DE density perturbation being given by (see Appendix~\ref{App:f(R)} for details)
\begin{align}\label{dRho_x}
3{\cal H}^2\Omega_x\, \delta_x =\;& 3{\cal H} \left(\Psi'+2{\cal H}\Phi\right)\dfrac{x_1 x_4}{1+x_4} - 3{\cal H}\delta{f}'_R \nn
& +\; 2\left[\nabla^2\Psi - 3{\cal H}(\Psi'+{\cal H}\Phi)\right] \dfrac{x_4}{1+x_4} \nn
& +\; \left[\nabla^2 + 3(x_3-1){\cal H}^2\right]\delta{f}_R ,
\end{align}
where $\delta_x \,{\equiv}\, \delta\rho_x/\bar{\rho}_x$ is the DE density contrast. 

By combining \eqref{dPsidt} and \eqref{grad2Psi}, we get the Poisson equation, given by
\bea\label{Poisson}
\nabla^2\Psi \;=\; 8\pi{G}a^2 \bar{\rho}_{\rm eff}\Delta_{\rm eff},
\eea
where $\Delta_{\rm eff} \,{=}\, \sum_A \Omega_A\Delta_A$ is effective comoving overdensity of the system, with the individual comoving overdensities being given by
\bea\label{Delta_A}
\bar{\rho}_A\Delta_A \;\equiv\; \delta\rho_A - 3{\cal H}\left(\bar{\rho}_A + \bar{p}_A\right)V_A,
\eea
where $\delta\rho_A$ is the coordinate density perturbation in the cosmic species, and given \eqref{V_x} and \eqref{dRho_x}, we have the DE comoving overdensity $\Delta_x$, given by 
\begin{align}\label{Delta_x}
3{\cal H}^2\Omega_x \Delta_x =\;& 3{\cal H}\left(\Psi'+{\cal H}\Phi\right) \dfrac{x_1 x_4}{1+x_4} + \dfrac{2x_4}{1+x_4}\nabla^2\Psi \nn
& +\; \left[\nabla^2+3\left(x_3-2\right){\cal H}^2\right]\delta{f}_R .
\end{align}

The Bardeen potentials are related according to the equation, given by
\bea\label{Phi_Psi}
\Psi - \Phi = 8\pi{G}a^2 \left(\bar{\rho}_{\rm eff} + \bar{p}_{\rm eff}\right)\Pi_{\rm eff} ,
\eea
where the effective anisotropic stress potential for the system is given by
\bea\label{Pi_eff}\nonumber
\Pi_{\rm eff} \;=\; \dfrac{1}{1+w_{\rm eff}} \sum_A\Omega_A \left(1 + w_A\right)\Pi_A,
\eea
with (see Appendix~\ref{App:f(R)} for details)
\bea\label{Pi_x}
3{\cal H}^2\Omega_x \left(1 + w_x\right) \Pi_x \;=\; \delta{f}_R + \left(\Psi-\Phi\right)\dfrac{x_4}{1+x_4}.
\eea
The expression for the DE anisotropic stress potential $\Pi_x$, is given in Appendix~\ref{App:f(R)}. Note that \eqref{Phi_Psi} may be used to eliminate the $\Phi$ terms from \eqref{d2Psidt2}---see Appendix~\ref{App:f(R)}. 

The off-diagonal field equations give the second-order evolution of the spatial Bardeen potential, given by (see Appendix~\ref{App:f(R)} for details)
\begin{align}\label{d2Psidt2}
&& \Psi'' + {\cal H} \left(2 + 3c^2_{a,\rm eff}\right) \Psi' \;+\;& {\cal H}\Phi' + (2x_3 + 3c^2_{a,\rm eff} - 1) {\cal H}^2\Phi \nn
&& +\; \dfrac{1}{3}\nabla^2\left(\Phi - \Psi\right) \;=\;& \dfrac{3}{2} {\cal H}^2 c^2_{s,\rm eff} \Delta_{\rm eff} ,
\end{align}
where the effective adiabatic sound speed of the system, is given by
\bea\label{c2a_eff}
c^2_{a,\rm eff} &=& \dfrac{1}{1+w_{\rm eff}} \sum_A \Omega_A\left(1+w_A\right)c^2_{aA}, \nn 
w_{\rm eff} &=& \sum_A \Omega_A w_A = \dfrac{1}{3}\left(1-2x_3\right), 
\eea
with $c^2_{aA} \,{\equiv}\, \bar{p}'_A/\bar{\rho}'_A$ being the square of adiabatic sound speed, and $w_{\rm eff} \,{\equiv}\, \bar{p}_{\rm eff}/\bar{\rho}_{\rm eff}$ is the effective equation of state parameter of the system, with $\bar{\rho}_{\rm eff}$, $\bar{p}_{\rm eff}$ and $x_3$ being given by \eqref{Friedmann}, \eqref{acceleration} and \eqref{x_i}, respectively, and $\Omega_A \,{=}\, \bar{\rho}_A/\bar{\rho}_{\rm eff}$ is the energy density parameter \eqref{Omega_A}; and the effective physical sound speed $c_{s,\rm eff}$, is given by
\bea\label{c2s_eff}
c^2_{s,\rm eff} = \dfrac{1}{\Delta_{\rm eff}} \sum_A \Omega_A \Delta_A c^2_{sA}, 
\eea
where $\Delta_{\rm eff}$ and $\Delta_A$ are as given by \eqref{Poisson} and \eqref{Delta_A}, respectively, and for standard fluids, the physical sound speed $c_{sA}$ is defined with respect to the rest frame of $A$.

The evolution equation of the perturbation $\delta{f}_R$, is 
\begin{align}\label{delfRPrime}
\delta{f}''_R + 2{\cal H}\delta{f}'_R =\;& \Big[ \nabla^2 + 2x_3 {\cal H}^2 \Big]\delta{f}_R + \dfrac{2{\cal H}^2F_x}{1+x_4} \Phi \nn
& +\; {\cal H} \left(3\Psi' + \Phi' + 4{\cal H}\Phi\right) \dfrac{x_1x_4}{1+x_4} \nn
& +\; {\cal H}^2 \Omega_m \Big[3\left(1-3c^2_{am} \right)(1 + w_m) {\cal H} V_m \nn
& +\; \left(1-3c^2_{sm} \right)\Delta_m \Big] - \dfrac{a^2\delta{R}}{3(1+x_4)} ,
\end{align}
where 
\bea
\delta{R} &=& -6a^{-2}\Big\{\Psi'' + 3{\cal H}\Psi' + {\cal H}\Phi' + 2x_3 {\cal H}^2\Phi \nn
&& \hspace{1.5cm} +\;  \dfrac{1}{3} \nabla^2\left(\Phi-2\Psi\right) \Big\} ,\\ \label{F_x}
F_x &{\equiv}& 1 -3w_m -2x_3 - x_4(1+3w_m)x_1 \nn
&& \hspace{0.2cm} -\; 3(1+w_m)(x_2+x_3)x_4,
\eea
with $\delta{R}$ being the perturbation in the Ricci scalar.


\subsection{The perturbed conservation equations}\label{subsec:consvEqs}
By the conservation of the total energy-momentum tensor \eqref{T_munu}, the perturbed balanced equations for any cosmic species $A$, may be given by
\begin{align}\label{ddelrhoAdt}
&& \delta{\rho}'_A +3{\cal H}\left(\delta{\rho}_A +\delta{p}_A\right) \;=\;& (\bar{\rho}_A +\bar{p}_A)\left[3\Psi' -\nabla^{2}V_A\right],\\
\label{dVAdt0}
&& \left[(\bar{\rho}_A+\bar{p}_A)V_A\right]' +\delta{p}_A \;=\;& -(\bar{\rho}_A +\bar{p}_A)\left[\Phi +4{\cal H}V_A\right] \nn
&& & -\; \frac{2}{3} \left(\bar{\rho}_A+\bar{p}_A\right) \nabla^{2}\Pi_A ,
\end{align}
and we may transform \eqref{ddelrhoAdt}, into 
\beq\label{ddAdt}
\delta_A' + 3{\cal H}\left(c^2_{sA} - w_A\right)\Delta_A - 3{\cal H} w_A'V_A = (1+w_A)\left[3\Psi' -\nabla^{2}V_A\right], 
\eeq 
where $\delta_A \,{\equiv}\, \delta{\rho}_A / \bar{\rho}_A$ is the coordinate overdensity or the density contrast, and we used 
\bea\label{dwAdt} 
w'_A = -3{\cal H}(1 \,{+}\, w_A)(c^2_{aA}-w_A),
\eea
with $w_A$ and $c_{aA}$ being as given in Secs.~\ref{subsec:backgrnd} and \ref{subsec:fieldEqs}, respectively. 

Instead of using \eqref{ddAdt}, it is rather important to consider the evolution of $\Delta_A$---given in \eqref{Delta_A}---as that is the density perturbation that appears in the Poisson equation \eqref{Poisson}, the $\Psi''$ equation \eqref{d2Psidt2}, and the physical sound speed of the system \eqref{c2s_eff}. Note that the Poison equation \eqref{Poisson} is a solution to the $\Psi'$ equation \eqref{dPsidt}. Thus, a common assumption in the literature, $\nabla^2\Psi \,{\approx}\, 4\pi{G}a^2\bar{\rho}_{\rm eff}\delta_{\rm eff}$, appears to be inconsistent, since this would not solve \eqref{dPsidt}. Moreover, for any fluid $A$, the quantity $\bar{\rho}_A\Delta_A$ corresponds to the density perturbation in the rest frame of that fluid \citep{Duniya:2015ths, Duniya:2015dpa}---and is unaffected by change of coordinates, unlike $\bar{\rho}_A\delta_A$ which is coordinate-dependent: consequently, $\delta_A$ is susceptible to (scale-dependent) gauge `artefacts' on very large scales.

Thus, the evolution of the velocity potential \eqref{dVAdt0} and the evolution of the density perturbation \eqref{ddAdt}, become
\begin{align} \label{dVAdt}
V_A' + {\cal H}V_A =& - \dfrac{1}{1+w_A} \left[c^{2}_{sA}\Delta_A + \dfrac{2}{3} \left(1+w_A\right) \nabla^{2}\Pi_A\right] \nn
&  -\Phi, \\
\label{dDAdt}
\Delta_A' -3{\cal H}w_A\Delta_A =&\; \dfrac{9}{2} {\cal H}^2 (1+w_A)\sum_B{\Omega_B(1+w_B)[V_A - V_B]} \nn
& -(1+w_A)\nabla^2 \left[V_A - 2{\cal H}\Pi_A\right], 
\end{align}
where we have used a general, non-adiabatic pressure perturbation, given by \citep{Duniya:2015ths, Duniya:2015dpa}
\bea\label{deltaPA}
\delta{p}_A = c^2_{aA}\delta\rho_A + \left(c^2_{sA}-c^2_{aA}\right)\bar{\rho}_A\Delta_A ,
\eea
with $\Delta_A$, $c_{aA}$ and $c_{sA}$ being as given in \eqref{Delta_A}, \eqref{c2a_eff} and \eqref{c2s_eff}, respectively. 

Thus, from \eqref{dVAdt} and \eqref{dDAdt}, the particular conservation equations are given as follows. The matter Euler and overdensity evolution equations, respectively, are given by
\bea\label{vmPrime}
V_m' + {\cal H}V_m &=& -\Phi , \\ \label{DmPrime}
\Delta_m' - \dfrac{9}{2} {\cal H}^2 \Omega_x(1+w_x)[V_m - V_x] &=& -\nabla^2V_m , 
\eea 
where henceforth, we assume pressureless matter ($\bar{p}_m = 0 = \delta{p}_m$); hence all pressure-dependent parameters vanish: $w_m \,{=}\, 0 \,{=}\, \Pi_m$ and $c_{am} \,{=}\, 0 \,{=}\, c_{sm}$. 

Note that we may use \eqref{V_x} and \eqref{Delta_x} for $V_x$ and $\Delta_x$, respectively---or alternatively, they may be solved for using \eqref{dVAdt} and \eqref{dDAdt}: by virtue of the definitions of the parameters in \eqref{rho_DE}, \eqref{p_DE}, \eqref{V_x}, \eqref{dRho_x} and \eqref{Pi_x}; with the pressure perturbation $\delta{p}_x$, being given by (see Appendix~\ref{App:f(R)} for details)
\begin{align}\label{delP_x}
3a^2{\cal H}^2\Omega_x w_x \dfrac{\delta{p}_x}{\bar{p}_x} =\;& \delta{f}''_R + {\cal H}\delta{f}'_R - 2\Phi f''_R \nn
& +\; \Big\{\dfrac{4}{3}\nabla^2 (\Phi-\Psi) + 2\left(\Psi'' + 2{\cal H} \Psi'\right) \nn
& +\; 2\left[ {\cal H}\Phi' + \left({\cal H}^2 + 2{\cal H}'\right) \Phi\right]\Big\} \dfrac{x_4}{1+x_4} \nn
& +\; {\cal H} \left[\Phi' + 2(\Psi' + {\cal H}\Phi)\right] \dfrac{x_1 x_4}{1+x_4} \nn
& -\; \left(\dfrac{4}{3}\nabla^2 +{\cal H}^2 + \dfrac{a^2}{6}\bar{R}\right)\delta{f}_R .
\end{align}
At this stage, let us stress that (in the given formalism) the $f(R)$ contribution corresponds to an “effective” fluid, having an energy-momentum tensor of the form given by \eqref{T_munu}; with the associated perturbations also obeying \eqref{ddelrhoAdt}--\eqref{deltaPA}.


\section{The Relativistic Magnification Overdensity}\label{sec:DeltaMag}
The observed, relativistic magnification overdensity \citep[see e.g.][]{Jeong:2011as, Duniya:2015ths, Duniya:2016ibg, Duniya:2016gcf, Duniya:2022xcz}, seen in a direction ${-}{\bf n}$ at a redshift $z$, is given by
\bea\label{Delta:obs} 
\Delta^{\rm obs}_{\cal M} ({\bf n},z)  = {\cal Q}(z)\, \hat{\delta}_{\cal M}({\bf n},z),
\eea 
where ${\cal Q}$ is the \emph{magnification bias} \citep{Blain:2001yf, Ziour:2008awn, Schmidt:2009rh, Schmidt:2010ex, Jeong:2011as, Liu:2013yna, Camera:2013fva, Duniya:2015ths, Duniya:2016ibg, Duniya:2015dpa, Duniya:2016gcf, Hildebrandt:2015kcb}, and $\hat{\delta}_{\cal M} \,{\equiv}\, \delta\hat{\cal M}/\bar{\cal M}$ is the magnification density contrast. Note that the {\em observed}, relativistic magnification overdensity \eqref{Delta:obs} is automatically {\em gauge-invariant}. In an inhomogeneous universe, objects get magnified or demagnified. 

The fact that we observe on the lightcone---and not on a spatial hypersurface---leads to the deformation of the image-plane surface, with the observation angles being distorted owing to weak (gravitational) lensing \citep[see e.g.][]{Schneider:2006eta, Bonvin:2008ni, Montanari:2015rga}. Apart from weak lensing, there are other sources of cosmic magnification. Time delay \citep[see e.g.][]{Raccanelli:2013gja, Baklanov:2020edg, Liao:2020yqz, Chan:2021jhh, Bayer:2021ugw, Baldwin:2021lud, Er:2022lad} also induce some distortion in the image plane. Moreover, by observing on the past lightcone, the observed $z$ becomes distorted by (i) Doppler effect \citep[see e.g.][]{Bonvin:2008ni, Bacon:2014uja, Raccanelli:2016avd, Bonvin:2016dze, Chen:2018hil, Andrianomena:2018aad, Coates:2020jzw}, by the motion of the sources relative to the observer, and (ii) the gravitational potential, both local at the sources (local potential-difference effects) and also integrated along the line of sight---integrated Sachs-Wolfe (ISW) effect \citep[see e.g.][]{LoVerde:2006cj, Raccanelli_2012, Ballardini:2018cho}. These effects, together with the time-delay effect, are otherwise known as \emph{relativistic effects}. Relativistic effects, with the exception of the Doppler effect, are mostly known to become significant at high $z \,{\gtrsim}\, 1$ and very large scales.

Thus, the image plane is distorted by lensing and relativistic effects, with the surface area per unit solid angle $\hat{\cal A}$ (in $z$ space) becoming (de)magnified by a factor $\mu \,{=}\, \hat{\cal M}/\bar{\cal M}$, given by \citep[e.g.][]{Duniya:2015ths, Duniya:2016ibg, Duniya:2016gcf}
\bea\label{Magnfcn}
\mu^{-1}({\bf n},z) \;\equiv\; \dfrac{\hat{\cal  A}({\bf n},z)}{\bar{\cal  A}(\bar{z})} \;=\; \dfrac{\hat{\cal D}^2_A({\bf n},z)}{\bar{\cal D}^2_A(\bar{z})},
\eea 
where $\bar{\cal  A}$ is the background part of the screen-space area density (i.e. the area density averaged over all solid angles); with $\hat{\cal M}$ being the magnification density (magnification per unit solid angle) in $z$ space, and $\hat{\cal D}_A$ is the associated angular diameter distance. Equation~\eqref{Magnfcn} implies that overdense regions will have a magnification factor $\mu \,{>}\, 1$ and objects appear closer than they actually are, with the apparent screen-space area appearing to be reduced or squashed. On the other hand, underdense regions will have $\mu \,{<}\, 1$ and objects tend to appear farther away, with the apparent screen-space area appearing stretched. Smooth, homogeneous regions will have $\mu \,{=}\, 1$ and objects are seen at their true position, with the apparent screen-space area remaining unchanged. Invariably, the apparent flux from an object becomes (de)amplified for ($\mu \,{<}\, 1$) $\mu \,{>}\, 1$; for $\mu \,{=}\, 1$, the apparent flux is the true flux.

\subsection{The image-plane area density}
Here we compute the image-plane area density $\hat{\cal A}$, being the area per unit solid angle transverse to the line of sight, at the image $z$. The transverse area element, is given by
\bea\label{AngDist}
dA = \hat{\cal A}({\bf n},z) d\Omega_{\bf n}, 
\eea 
where the area density $\hat{\cal A} \,{=}\, \partial A/\partial \Omega_{\bf n}$ is as given in \eqref{Magnfcn}, with $\Omega_{\bf n}$ being the solid angle along the direction ${-}{\bf n}$. Given the spacetime metric~\eqref{metric1}, we consider the conformal transformation, given by
\bea\label{metric-trans}
ds^2 \to d\tilde{s}^2 &=& a^2 ds^2, \nn
&=& a^2\left\{-(1+2\phi) d\eta^2 + 2B_i dx^i d\eta \right.\nn
&& +\; \left. \left[(1-2\psi)\delta_{ij} + 2E_{|ij}\right] dx^i dx^j \right\},
\eea
where $\psi \equiv D+\frac{1}{3}\nabla^2 E$, and $E_{|ij} \equiv \nabla_i\nabla_j E$. Then in real-space coordinates $\tilde{x}^\alpha$, which correspond to the conformal metric $d\tilde{s}$, we have
\bea\label{dA1}
dA &=& \sqrt{-\tilde{g}}\, \epsilon_{\mu\nu\alpha\beta}\, \tilde{u}^\mu \tilde{\ell}^\nu d\tilde{x}^\alpha d\tilde{x}^\beta,\nn \label{dA2}
&\;\equiv\; & {\cal A} (\theta_O,\vartheta_O)\, d\theta_O d\vartheta_O, 
\eea
where $\theta_O$ and $\vartheta_O$ being the zenith and the azimuthal angles, respectively, at the observer $O$, and $\tilde{u}^\nu$ is the 4-velocity of the observer. The 4-vector $\tilde{\ell}^\nu \,{=}\, \tilde{u}^\nu + \tilde{n}^\nu/(\tilde{n}^\alpha \tilde{u}_\alpha)$ \citep{Jeong:2011as, Duniya:2015ths, Duniya:2016ibg, Duniya:2016gcf} lies in the image plane and hence is orthogonal to the line of sight, i.e.~$\tilde{u}_\nu\tilde{\ell}^\nu \,{=}\, 0$, with $\tilde{n}^\nu \,{=}\, d\tilde{x}^\nu / d\lambda$ being a tangent 4-vector to the photon geodesic $\tilde{x}^\nu(\lambda)$ and, $\lambda$ being an affine parameter. 

Note that $\hat{\cal A}$ and ${\cal A}$, in \eqref{AngDist} and \eqref{dA2}, are the area densities in $z$ space and in real space, respectively. From \eqref{dA2}, we have the real-space area density, given by
\bea\label{A:defn}
{\cal A} = \sqrt{-\tilde{g}}\, \epsilon_{\mu\nu\alpha\beta}\, \tilde{u}^\mu \tilde{\ell}^\nu \dfrac{\partial\tilde{x}^\alpha}{\partial\theta_S} \dfrac{\partial\tilde{x}^\beta}{\partial\vartheta_S} \left| \left. \dfrac{\partial(\theta_S,\vartheta_S)}{\partial(\theta_O,\vartheta_O)} \right| \right. ,
\eea
where $\tilde{g} \,{=}\, {\rm det}(\tilde{g}_{\mu\nu})$, with $\theta_S \,{=}\, \theta_O + \delta\theta$ and $\vartheta_S \,{=}\, \vartheta_O+\delta\vartheta$ being the angles at the source $S$. Thus, after some calculations \citep[see e.g.][for details]{Duniya:2015ths, Duniya:2016ibg, Duniya:2016gcf}, we have 
\bea\label{A:exprssn}
\dfrac{{\cal A}}{\bar{\cal A}} &=& 1 - 3D - \phi + \bar{n}^i B_{|i} - \dfrac{1}{2} \delta{g}_{\alpha\beta}\bar{n}^\alpha \bar{n}^\beta + 2\dfrac{\delta{r}}{\bar{r}} \nn
&& +\; \left(\cot{\theta} +\partial_{\theta}\right)\delta{\theta} +\partial_{\vartheta}\delta{\vartheta},\nn \label{A:exprssn2}
&=& 1 - 2\Psi + \int^{\bar{r}_S}_0{ d\bar{r}\left[\dfrac{2}{\bar{r}_S} - \left(\bar{r}_S - \bar{r}\right) \dfrac{\bar{r}}{\bar{r}_S } \nabla^2_\perp \right] \left(\Phi + \Psi\right) } \nn
&& \quad +\; 2\left(1 - \dfrac{1}{\bar{r}_S {\cal H}}\right) {\cal H}\left(E'-B\right),
\eea
where $\bar{\cal A} \,{=}\, a^2\bar{r}^2\sin{\theta}_O$ is the background area density, $r \,{=}\, \bar{r} + \delta{r}$ is the comoving radial distance, with $\bar{r}_S \,{=}\, \bar{r}(\bar{z}_S)$ being the background comoving distance at $S$; the rest of the parameters are as given in Sec.~\ref{sec:fR}, with $\nabla^2_\perp \,{=}\, \nabla^2 - (\bar{n}^i\partial_i)^2 + 2\bar{r}^{-1} \bar{n}^i\partial_i$ being the Laplacian (the various terms retaining their standard notations) on the image plane, transverse to the line of sight. (The various terms in $\nabla^2_\perp$ retain their standard definitations.)

\subsection{The magnification distortion} 
Here we relate the $z$-space area density $\hat{\cal A}$ to the real-space area density ${\cal A}$. By taking a gauge transformation from real to $z$ space, in first-order perturbations, we have
\bea\label{perp_dA}
\hat{\cal A}({\bf n},z) = {\cal A}({\bf n},z) -\dfrac{d\bar{\cal A}}{d\bar{z}} \delta{z}({\bf n},z),
\eea
where here ${\cal A}$ is expressed as a function of $z$, and
\bea
\dfrac{d\bar{\cal A}}{d\bar{z}} = -2\left(1 - \dfrac{1}{\bar{r} {\cal H}}\right) \dfrac{\bar{\cal A}}{1+\bar{z}} ,
\eea
noting that $a \,{=}\, 1/(1+\bar{z})$; the $z$ perturbation is given by 
\bea\label{deltaz}
\dfrac{\delta{z}}{1+\bar{z}} = - \Big[\Phi + \Psi + {\bf n}\,{\cdot}{\bf V} - \psi\Big]^S_O - \int^{\bar{r}_S}_0 d\bar{r} \left(\Phi' +\Psi'\right),
\eea
where the scalar field $\psi$ is as given by \eqref{metric-trans}, and ${\bf V} \,{=}\, \partial_i V$ is the gauge-invariant velocity, with $V$ being the velocity potential as given by \eqref{DefnV}.

By combining \eqref{A:exprssn2}--\eqref{deltaz}, and using $\mu^{-1} = 1 - \hat{\delta}_{\cal M} = \hat{\cal A}/\bar{\cal A}$, we get the magnification distortion $\tilde{\delta}_{\cal M}$ in a relativistic form. Consequently, we have the (observed) relativistic magnification overdensity \eqref{Delta:obs}, given by
\begin{align} \label{Delta:obs2}
\Delta^{\rm obs}_{\cal M} =&\; {\cal Q}\left[1 - \dfrac{\hat{\cal A}({\bf n},z)}{\bar{\cal A}(\bar{z})} \right],\\ \label{Delta:obs3}
=&\; - {\cal Q}\int^{\bar{r}_S }_0{ d\bar{r}\left[\dfrac{2}{\bar{r}_S} - \left(\bar{r}_S - \bar{r}\right) \dfrac{\bar{r}}{\bar{r}_S } \nabla^2_\perp \right] \left(\Phi + \Psi\right) } \nn
& +\; 2{\cal Q}\left(1 -\dfrac{1}{\bar{r}_S {\cal H}}\right) \left[ \Phi + {\bf n}\,{\cdot}{\bf V} + \int^{\bar{r}_S }_0{d\bar{r} \left(\Phi' + \Psi' \right)} \right] \nn
& +\; 2{\cal Q}\Psi ,
\end{align}
where non-integral terms denote relative values, i.e.~the values at $S$ relative to those at $O$. \citep[See][for the full details of \eqref{perp_dA}--\eqref{Delta:obs3}.]{Duniya:2015ths, Duniya:2016ibg, Duniya:2016gcf} 

For the rest of this work, we take the observed magnification overdensity to be given by
\bea\label{DeltaObs_M} 
\Delta^{\rm obs}_{\cal M}({\bf n},z) = \Delta^{\rm std}_{\cal M}({\bf n},z) + \Delta^{\rm rels}_{\cal M}({\bf n},z),
\eea
where we take the weak (gravitational) lensing magnification term as the {\em standard} component, given by
\bea\label{MagStd} 
\Delta^{\rm std}_{\cal M} \;\equiv\; -{\cal Q} \int^{\bar{r}_S }_0{ d\bar{r}\left(\bar{r} - \bar{r}_S \right) \dfrac{\bar{r}}{\bar{r}_S} \nabla^2_\perp \left(\Phi + \Psi\right)},
\eea
and the relativistic-correction component, is given given by 
\begin{align}\label{MagGR}
\Delta^{\rm rels}_{\cal M} \;\equiv\;& 2{\cal Q} \left\lbrace \left(1 -\dfrac{1}{\bar{r}_S {\cal H}}\right) \left[ \Phi + {\bf n}\,{\cdot} {\bf V}  + \int^{\bar{r}_S}_0{d\bar{r} \left(\Phi' + \Psi' \right) } \right] \right. \nn
& +\; \Psi - \left. \dfrac{1}{\bar{r}_S} \int^{\bar{r}_S}_0{ d\bar{r} \left(\Phi + \Psi\right)} \right\rbrace .
\end{align}
As given by \eqref{DeltaObs_M}, we have that apart from weak lensing \eqref{MagStd}, the cosmic magnification is also sourced by relativistic effects \eqref{MagGR}: the Doppler effect term (velocity potential term in square brackets); the ISW-effect term (integral term in square brackets); the time-delay effect term (last integral term), and the line-of-sight gravitational potential-well term (non-integral potential terms).


\section{The Magnification Angular Power Spectrum}\label{sec:MagCls}%
The observed magnification overdensity \eqref{Delta:obs3} is expanded in spherical multipoles, given by
\bea\label{Delta:ell}
\Delta^{\rm obs}_{\cal M}({\bf n},z) &=& \sum_{\ell m} {a_{\ell m}(z) Y_{\ell m}({\bf n})}, \nn 
a_{\ell m}(z) &=& \int{ d^2{\bf n}\, Y^*_{\ell m}({\bf n}) \Delta^{\rm obs}_{\cal M}({\bf n},z) },
\eea
where $Y_{\ell m}({\bf n})$ are the spherical harmonics and $a_{\ell m}$ are the multipole expansion coefficients, with the asterisk denoting complex conjugate. The angular power spectrum observed at a source redshift $z_S$, is given by 
\begin{align}\nonumber 
C_\ell(z_S) =& \left\langle{\left.\left| a_{\ell m}(z_S) \right.\right|^2 }\right\rangle, \\ \label{Cl_TT2}
=& \dfrac{4}{\pi^2} \left(\dfrac{43}{50}\right)^2 \int{dk\, k^2 T(k)^2 P_{\Phi_p}(k) \Big|f_\ell(k,z_S) \Big|^2 },
\end{align}
where $k$ is the wavenumber, and we have 
\begin{align}\label{f_ell} 
f_\ell(k,z_S) =&\; {\cal Q} \int^{\bar{r}_S}_0{d\bar{r}\, \ell(\ell+1) j_\ell(k\bar{r}) \dfrac{(\bar{r}-\bar{r}_S)}{\bar{r}_S\bar{r}}  \left(\check{\Phi}+\check{\Psi}\right) (k,\bar{r})} \nn
& -2{\cal Q} \left(1 - \dfrac{1}{\bar{r}_S {\cal H}}\right) \Big\{ \check{V}^\parallel_m(k,z_S) \partial_{k\bar{r}} j_\ell(k\bar{r}_S) \nn
& \hspace{1.3cm} - \int^{\bar{r}_S}_0{d\bar{r}\, j_\ell(k\bar{r}) \left(\check{\Phi}' + \check{\Psi}'\right) (k,\bar{r})} \nn
& \hspace{1.3cm} -\; \left(1 - \dfrac{1}{\bar{r}_S {\cal H}}\right) j_\ell(k\bar{r}_S) \check{\Phi}(k,z_S) \nn
& \hspace{1.3cm} -\; j_\ell(k\bar{r}_S) \check{\Psi}(k,z_S) \Big\rbrace \nn
& -\dfrac{2{\cal Q}}{\bar{r}_S} \int^{\bar{r}_S}_0 d\bar{r}\, j_\ell(k\bar{r}) \left(\check{\Phi}+\check{\Psi}\right),
\end{align}
where $V^\parallel_m \,{=}\, {-}n^i\partial_i V_m$ is the line-of-sight matter peculiar velocity of the source relative to the observer (assuming that on large scales, being the scales considered in this work, galaxies trace the same trajectories as the underlying matter), and $V_m$ is the gauge-invariant matter velocity potential. We use $\partial_{kr} \,{=}\, \partial / \partial(kr)$, and $j_\ell$ is the spherical Bessel function; we use the notation $\check{X}(k,z) \equiv X(k,z)/\Phi_d(k)$ for a given parameter $X$, in \eqref{f_ell}, with~\citep{Duniya:2013eta, Duniya:2015nva, Duniya:2015ths, Duniya:2019mpr}
\bea\label{Phi_d}
\Phi(k,z_d) \;=\; \dfrac{43}{50} \Phi_p(k) T(k) \;\equiv\; \Phi_d(k),
\eea
where $\Phi_d$ gives the gravitational potential at the epoch of photon-matter decoupling $z \,{=}\, z_d$, $\Phi_p$ is the primordial gravitational potential; $\check{X}$ essentially measures the growth function of the associated parameter.

\subsection{The $f(R)$ gravity model}\label{subsec:model}
In this work we only consider the well known Hu-Sawicki model, given by \citep{Hu:2007nk, Tsujikawa:2007xu, Amendola:2010bk, Clifton:2011jh, Ishak:2018his}
\bea\label{HuSawicki}
f(R) = -\lambda R_c \dfrac{\left(R/R_c\right)^n}{1+\left(R/R_c\right)^n},
\eea
where $n$, $\lambda$ and $R_c$ are the degrees of freedom of the model; $R$ is the Ricci scalar (see Sec.~\ref{sec:fR}). We see that for all values $n \,{\neq}\, 0$, the function $f(R)$ is nonlinear in $R$. The exponent $n$ dictates the strength of the gravity, so that the larger the its value the stronger the gravity.

For the purpose of our analysis, we adopt (henceforth) a matter density parameter $\Omega_{m0} \,{=}\, 0.24$, a Hubble constant $H_0 \,{=}\, 73 \kms\cdot\mpc$, and a constant magnification bias, ${\cal Q}(z) \,{=}\, 1$. We also use a DE physical sound speed $c_{sx} \,{=}\, 1$ \citep[the value from quintessence][]{Amendola:2010bk, Duniya:2013eta}, an adiabatic sound speed $c_{ax}$ as given by \eqref{c2a_eff}; with pseudo-$\Lambda$CDM initial conditions for the background, and adiabatic initial conditions for the perturbations (see Appendix~\ref{App:ICs}, for the initial conditions). For all numerical calculations we set $\lambda \,{\simeq}\, 1.328$ and $R_c \,{=}\, 18.75\, \Omega_{m0}\, H_0^2$, and we initialize all evolutions in the matter domination epoch.


\subsection{The imprint of $f(R)$ gravity}\label{subsec:imprint}
As previously stated, the main goal of this investigation is to provide a qualitative analysis of the imprint of $f(R)$ gravity in the cosmic magnification, on large scales, i.e. near and beyond the Hubble radius. Our approach focuses on the effects arising primarily from the perturbations. Thus, we set the background cosmological expansion history to be identical for all the (different) values of the given gravity parameters, at all $z$. The advantage of this, is that all deviations from standard cosmology are both restricted to the perturbations and also isolated on the largest scales.

\begin{figure}\centering
\includegraphics[width=8.5cm,height=12cm]{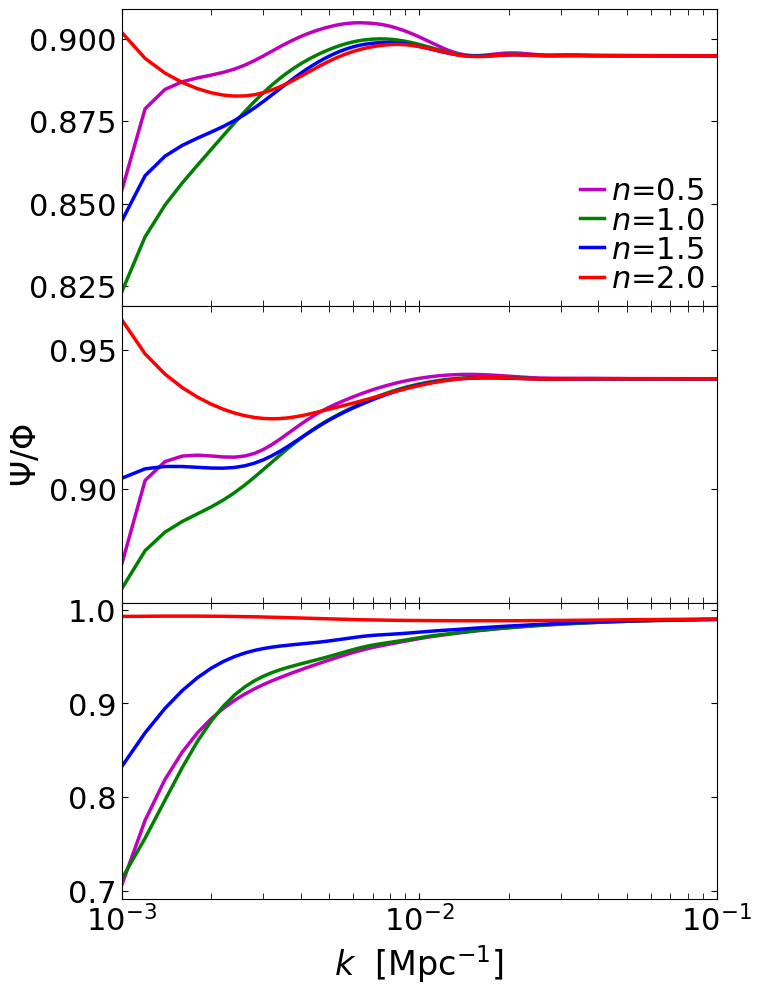} 
\caption{The ratio of the spatial to the temporal metric potentials, $\Psi$ and $\Phi$, respectively, with respect to the wavenumber $k$; for the values of the exponent $n \,{=}\, 0.5$ (magenta), $1.0$ (green), $1.5$ (blue), and $2.0$ (red), at source redshifts $z_S \,{=}\, 0.5$ (top panel), $z_S \,{=}\, 1$ (middle panel), and $z_S \,{=}\, 3$ (bottom panel). In $\Lambda$CDM, $\Psi(k)/\Phi(k) \,{=}\, 1$, at all $z$.}\label{Psi2Phiratios}
\end{figure}

\begin{figure*}\centering
\includegraphics[width=16cm,height=12cm]{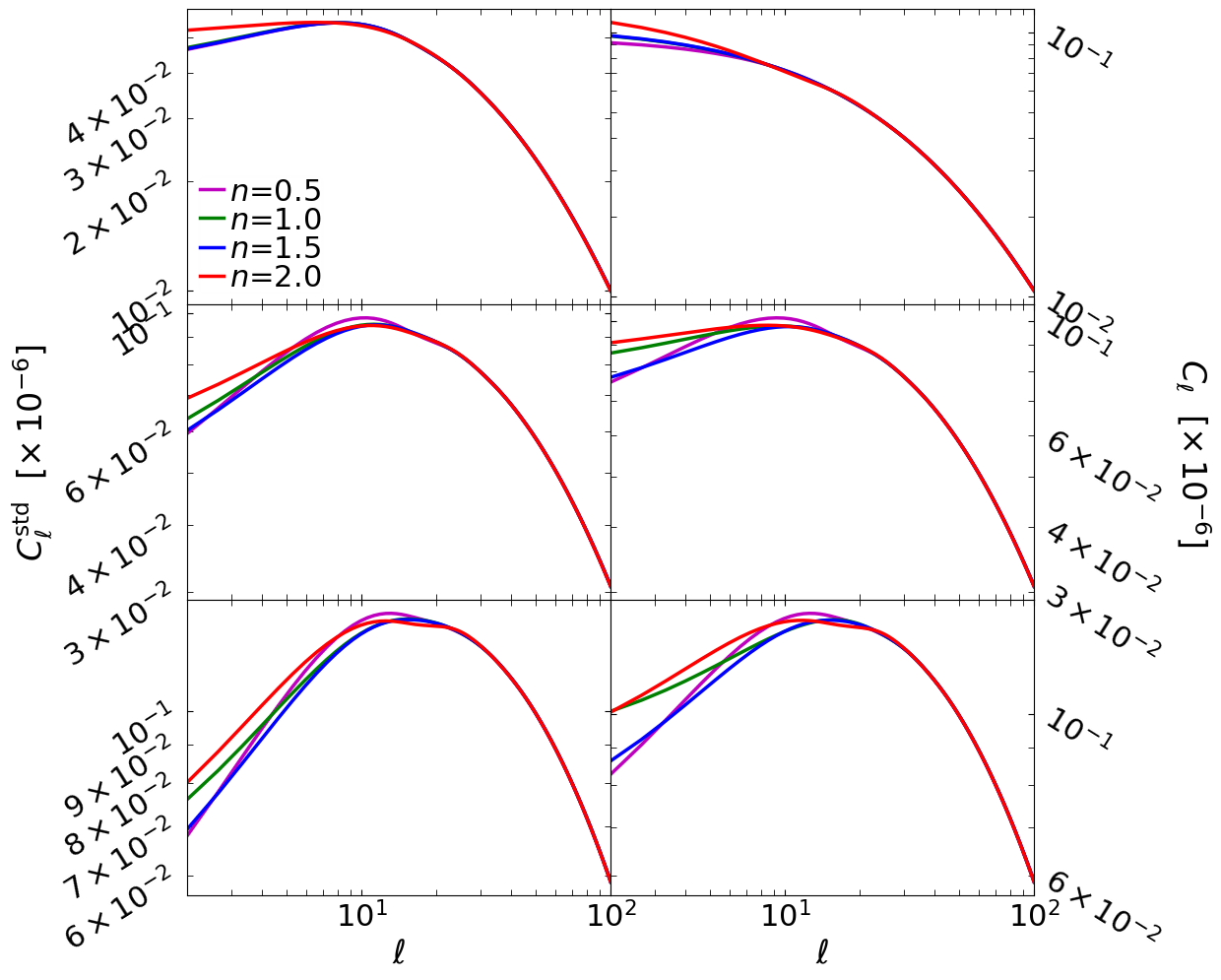} 
\caption{\emph{Left panels:} The standard (lensing) magnification power spectrum  $C^{\rm std}_\ell$, with respect to multipoles $\ell$. \emph{Right panels:} The total (relativistic) magnification angular power spectrum $C_\ell$, with respect to $\ell$. Both $C^{\rm std}_\ell$ and $C_\ell$ are given for the values of $n \,{=}\, 0.5$ (magenta), $1.0$ (green), $1.5$ (blue), and $2.0$ (red), at source redshifts $z_S \,{=}\, 0.5$ (top panels), $z_S \,{=}\, 1$ (middle panels), and $z_S \,{=}\, 3$ (bottom panels).}\label{stdtotCls}
\end{figure*}

In Fig.~\ref{Psi2Phiratios}, we show the plots of the ratio $\Psi / \Phi$ as a function of wavenumber $k$ for the values of the model parameter $n \,{=}\, 0.5,\, 1,\, 1.5,$ and $2$, at source redshifts $z_S \,{=}\, 0.5,\, 1,$ and $3$ (top to bottom, respectively). Given the fact that in the cos-
mological concordance model ($\Lambda$CDM) this ratio is an absolute unity (on all scales and at all z), this ratio will measure any changes in the perturbations owing to the $f(R)$ gravity.

Thus, in Fig.~\ref{Psi2Phiratios} we see that on smaller scales (larger $k$) the ratios converge on a single value and remain so at all $z$. On larger scales (smaller $k$), we observe the ratios for the different values of $n$ deviating from the common (fixed) value obtained on the small scales. These changes are a consequence of our normalization: marching all evolutions to the same background universe. Thus, these results measure solely the large-scale imprint of the underlying gravity (with no background effects). We see that at $z \,{<}\, 3$, there is an inconsistent behaviour in the ratios for $n \,{=}\, 1.5$ and $n \,{=}\, 2$, by these ratios crossing over the others; whereas, at $z \,{\geq}\, 3$, we see a consistent behaviour in the ratios for all the values of $n$, with the ratio for $n \,{=}\, 2$ approaching unity (the value for $\Lambda$CDM). This implies that at $z \,{\geq}\, 3$, the two potentials in the given gravity model will follow similar evolutionary tracks for $n \,{\simeq}\, 2$. Moreover, the large-scale behaviour exhibited by the ratios for $n \,{=}\, 1.5$ and $n \,{=}\, 2$ at $z \,{<}\, 3$, may be an indication that values of $n \,{\gtrsim}\, 1.5$ (or $n \,{>}\, 1$) are not admissible by the given $f(R)$ model with respect to cosmological perturbations. These values may not be physical or realistic at the given $z$, i.e. they may be too large to produce a physically meaningful cosmology of the large scale structure, despite being capable of producing a viable background behaiour. (Further analysis which involve a quantitative approach may be required to assertain this.) 

In Fig.~\ref{stdtotCls}, left column, we present the plots of the standard (lensing) magnifiaction angular power spectrum $C^{\rm std}_\ell$ for the chosen values of $n \,{=}\, 0.5,\, 1,\, 1.5,$ and $2$, at source redshifts $z_S \,{=}\, 0.5,\, 1,$ and $3$ (top to bottom, respectively). We see that the plots (on their own) do not seem to exhibit a clear, particular behaviour or pattern for the different values of $n$, at the given $z_S$. However, we see that as $z$ increases, the ratios become more separated on larger scales (smaller $\ell$), indicating that the effect of $n$ gradually becomes more prominent as we move toward earlier epochs. This is understandable since lensing grows as $z$ increases: as given in Sec.~\ref{sec:DeltaMag}, lensing phenomenon in the large scale structure is an integral effect (which is prescribed by the observed $z$ interval). Moreover, we see that the plots for the different values of $n$ match each other on smaller scales (larger $\ell$), which can be understood as a consequence of our normalization.

Also in Fig.~\ref{stdtotCls}, right column, we present the plots of the total (relativistic) magnifiaction angular power spectrum $C_\ell$, given by \eqref{Cl_TT2}, for the same parameters as for $C^{\rm std}_\ell$ (left column). As expected, we see that the angular power spectra for the different values of $n$ coincide on smaller scales (larger $\ell$), at all $z$; however, they deviate on the largest scales. We also observe that the value of $\ell$ at which the deviations begin, increases with $z$; this value of $\ell$ also appears to coincide with the position of the onset of the turnover in the magnification angular power spectrum. We see that the higher the $z$, the smaller the value of $\ell$ at which the deviations begin, and the more prominent the turnover in the angular power spectrum. Conversely, as we move toward the present epoch ($z \,{=}\, 0$), the deviations gradually converge to a single result (similar to the $C^{\rm std}_\ell$ scenario). 

Furthermore, there seems to be a subtle pattern in the large-scale separation of the lines of $C_\ell$ for the different values of $n$: as $z$ increases, the lines for integral values of $n$ appear to group together, with relatively larger amplitude, while the lines for decimal values of $n$ also appear to group together separately, with a relatively lower amplitude. This feature can be searched for in quantitative analyses in the experimental data---which could serve as a ``smoking gun" for this $f(R)$ gravity model. (Future high-precision surveys should be able to detect elusive deviations from standard cosmology, in the large-scale structure.)

Moreover, given that full relativistic effects were taken into account in $C_\ell$, apparently they are responsible for the large-scale changes (or effects) observed in $C_\ell$, relative to those in $C^{\rm std}_\ell$. In fact, apart from the subtle grouping pattern previously highlighted, we clearly see that the separation between these two groups of lines tend to increase with increasing $z$, an effect which is barely noticeable in the $C^{\rm std}_\ell$ scenario. Thus, including the relativistic corrections in the magnification overdensity could help enhance the potential of the cosmic magnification as a cosmological probe.

\begin{figure}\centering
\includegraphics[width=8.5cm,height=12cm]{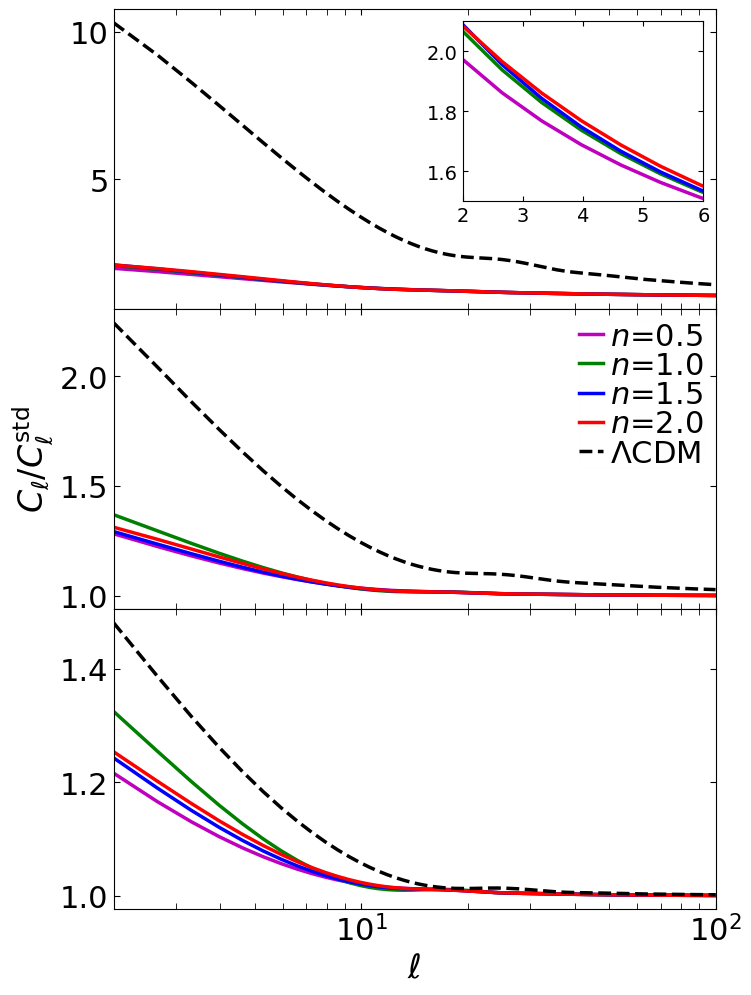} 
\caption{The ratio of the total (relativistic) magnification angular power spectrum $C_\ell$ to the standard (lensing) magnification angular power spectrum $C^{\rm std}_\ell$, with respect to multipoles $\ell$, for the values of $n \,{=}\, 0.5$ (magenta), $1.0$ (green), $1.5$ (blue), and $2.0$ (red), at source redshifts $z_S \,{=}\, 0.5$ (top panel), $z_S \,{=}\, 1$ (middle panel), and $z_S \,{=}\, 3$ (bottom panel). The black dashed line gives the corresponding ratio in $\Lambda$CDM.}\label{tot2stdratios}
\end{figure}

In Fig.~\ref{tot2stdratios} we show the ratio of the total (relativistic) magnification angular power spectrum $C_\ell$ to the standard manginifaction angular power spectrum $C^{\rm std}_\ell$ as a function of $\ell$, for the same $f(R)$ parameters as in Figs.~\ref{Psi2Phiratios} and~\ref{stdtotCls}. In this figure we have also shown the ratio for $\Lambda$CDM. These ratios measure the total contribution of relativistic effects in the magnification angular power spectrum. We see that the amplitude of the ratio for $\Lambda$CDM is larger than those of $f(R)$ gravity for all the values of $n$, on the largest scales. However, we see that as $z$ increases, the difference in amplitude between the two models gradually reduces; with the amplitudes becoming of the same order of magnitude at $z \,{\gtrsim}\, 3$. Thus, at $z \,{<}\, 3$, cosmic magnification in $\Lambda$CDM will have larger relativistic effects than in $f(R)$ gravity. This is not surprising since there is relatively stronger curvature in $f(R)$ at lower $z$ (towards $z \,{=}\, 0$); hence the amplitude of the gravitational potentials (and matter) are relatively more suppressed in $f(R)$ gravity, at the given $z$ (see amplitudes of $\Psi/\Phi$ in Fig.~\ref{Psi2Phiratios}). Consequently, we have a relatively diminished magnification angular power spectrum, and invariably, relativistic effects. At higher $z$, in the matter domination epoch, both $\Lambda$CDM and $f(R)$ have similar evolution in the perturbations; hence giving similar relativistic effects. Moreover, similar to results in Figs.~\ref{Psi2Phiratios} and~\ref{stdtotCls}, we see that as $z$ increases, the ratios of the magnification angular power spectrum in $f(R)$ gravity exhibit an inconsistent behaviour for values of $n \,{\gtrsim}\, 1.5$. This may be an indication that $n \,{>}\, 1$ are not admissible by the given $f(R)$ model. (In order to understand this better, further analysis may be needed, which is outside the scope of this work.)

\begin{figure*}\centering
\includegraphics[width=16cm,height=12cm]{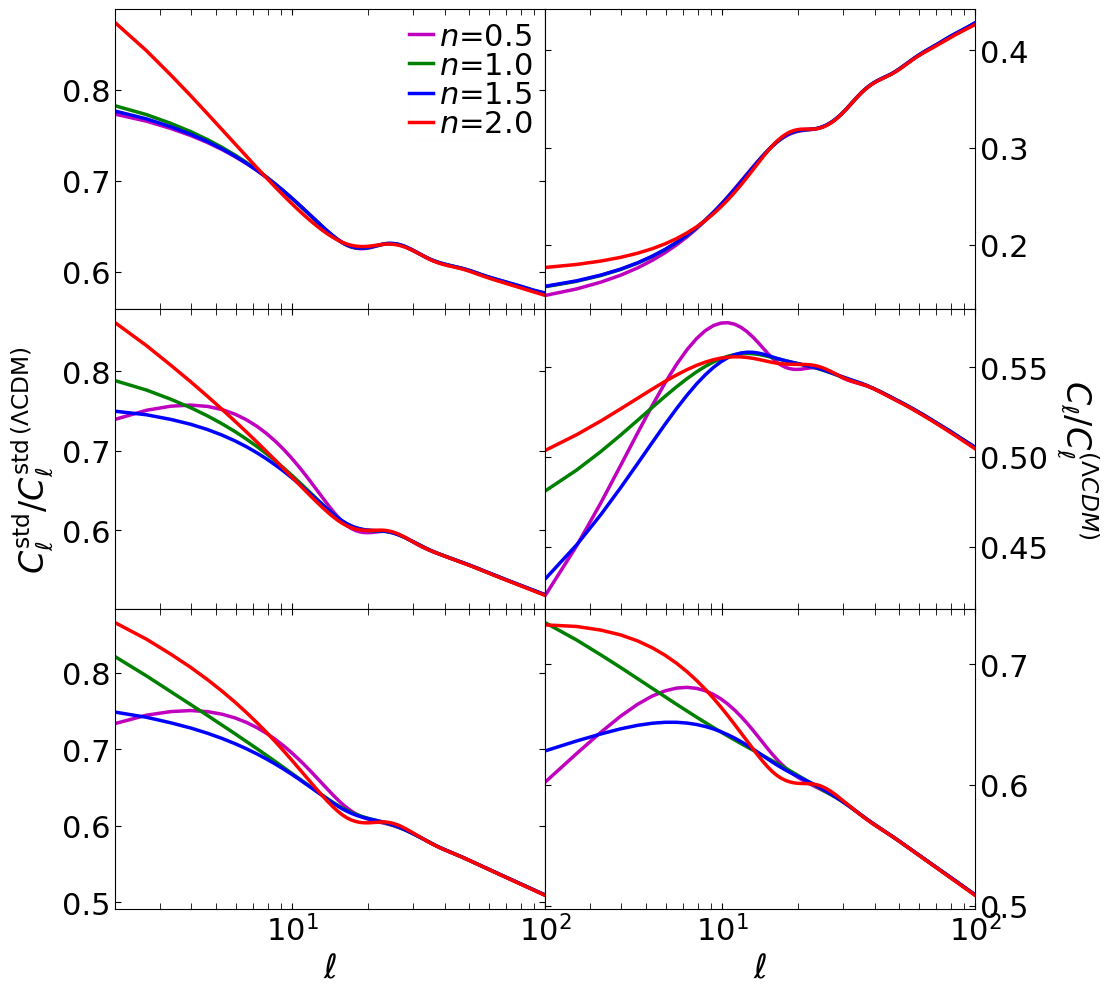} 
\caption{\emph{Left panels:} The ratios of the standard (lensing) magnification power spectrum $C^{\rm std}_\ell$ in $f(R)$ gravity to that in $\Lambda$CDM, with respect to multipoles $\ell$. \emph{Right panels:} The ratios of the total (relativistic) magnification angular power spectrum $C_\ell$ in $f(R)$ gravity to that in $\Lambda$CDM, with respect to multipoles $\ell$. Both $C^{\rm std}_\ell$ and $C_\ell$ ratios are given for the values of $n \,{=}\, 0.5$ (magenta), $1.0$ (green), $1.5$ (blue), and $2.0$ (red), at source redshifts $z_S \,{=}\, 0.5$ (top panels), $z_S \,{=}\, 1$ (middle panels), and $z_S \,{=}\, 3$ (bottom panels).}\label{fofR2LCDMratios}
\end{figure*}

Finally, in Fig.~\ref{fofR2LCDMratios}, left column, we compare the standard (lensing) magnification angular power spectrum $C^{\rm std}_\ell$ in $f(R)$ gravity (for the same parameters as in Figs.~\ref{Psi2Phiratios}--\ref{tot2stdratios}) to that in $\Lambda$CDM. The observed behaviour in the plots follow mostly from the previous figures. It should be pointed out that the behaviour of these ratios is not only owing to changes in the perturbations (unlike in Figs.~\ref{Psi2Phiratios}--\ref{tot2stdratios}), but also owing to the background cosmology (which is markedly different for $f(R)$ gravity and $\Lambda$CDM.) However, since the angular power spectra are computed at fixed source redshifts, the background will mainly provide a constant contribution (or effect) in the angular power spectrum. Thus, we see that at all the source redshifts, although the various ratios converge to a single value on smaller scales (larger $\ell$), this value is different from unity: it measures the background effect, in $f(R)$ gravity relativie to $\Lambda$CDM, at the given $z_S$. We also observe that, although the amplitude of the ratios is less than unity, at all the values of $z_S$ and $\ell$, the ratios themselves appear to be growing increasingly on the largest scales. This implies that on the largest scales, the amplitude of the lensing magnification angular power spectrum in $\Lambda$CDM decreases relatively quickly (with its turnover becoming more prominent as $z$ increases) towards the value of the amplitude of the angular power spectrum in $f(R)$ gravity.

Similarly in Fig.~\ref{fofR2LCDMratios}, right column, we compare the total (relativistic) magnification angular power spectrum $C_\ell$ in $f(R)$ gravity (for the same parameters as in Figs.~\ref{Psi2Phiratios}--\ref{tot2stdratios}) to that in $\Lambda$CDM. In general, similar discussion follows as for the results in the left column, except that unlike in the left column where we observe the ratios of the lensing manification angular power spectra to be growing increasingly on the largest scales at all the values of $z_S$, here (with the relativistic corrections included in the observed magnification overdensity) we see that at $z_S \,{\leq}\, 1$ the ratios are rather continuously decreasing on the largest scales. This implies that at the given values of $z_S$, the amplitude of the relativistic magnification angular power spectrum in $f(R)$ decreases quickly (with decreasing $z$) relativie to that of $\Lambda$CDM. This suggests that the combined relativistic corrections in the cosmic magnification angular power spectrum in $f(R)$ gravity have a net negative effect (or contribution) which diminishes the large-scale angular power, at the given $z$; whereas at higher $z \,{>}\, 1$, the relativistic corrections combine to give a net positive effect---thereby boosting the amplitude of the total magnification angular power spectrum in $f(R)$ relative to that in $\Lambda$CDM. 

Moreover, similar to the results in Fig.~\ref{stdtotCls} (right column), we see that as $z$ increases, the ratios of the total magnification angular power spectra exhibit a subtle pattern in which the ratios for integral values of $n$ tend to cluster together, and those for decimal values also tend to cluster together (separately). This feature could be searched for in the experimental data, which may serve as a signature for the given $f(R)$ model in the cosmic magnification, at higher $z$.


\section{Conclusion}\label{sec:Conc}
We presented a qualitative investigation of the effects of $f(R)$ gravity on the cosmic magnification, on large scales, using the Hu-Sawicki $f(R)$ model. To achieve this, we normalised all perturbations evolutions to the same Robertson-Walker (spatially) flat background spacetime, at all redshifts $z$. This ensures that all deviations from standard, $\Lambda$CDM cosmology, observed in the given $f(R)$ gravity model are restricted to the perturbations and isolated on the largest scales (at all $z$). We also took care to include all the known relativistic corrections to the magnification overdensity.

We compared the spatial (spacetime) metric potential to the temporal metric potential in $f(R)$ gravity by taking the ratio of the two parameters, with respect to $k$ (wavenumber), for four values of the the gravity parameter $n$ ($0.5 \,{\leq}\, n \,{\leq}\, 2$) in the Hu-Sawicki $f(R)$ model, at different source redshifts $z_S$. As expected from our normalisation, the ratios coincide on smaller scales, at all $z$; whereas on larger scales, the ratios deviate (differently) from the small-scale common value. Moreover, at $z \,{<}\, 3$, there appeared to be irregular changes on the largest scales for values of $n \,{\gtrsim}\, 1.5$: the lines of the ratios cross over those of lower values of $n$. This irregular behaviour may be an indication that values of $n \,{>}\, 1$ may not be admissible in the Hu-Sawicki model, with respect to the cosmological perturbations.

We also computed the cosmic magnifiaction angular power spectrum for the chosen values of $n$ and $z_S$. The results showed that the changes induced by $f(R)$ gravity in the cosmic magnification are enhanced by relativistic effects, as $z$ increases. Relativistic effects cause the angular power spectra for the given values of $n$ to become more prominent, and they also induce a subtle pattern in the separation of the angular power spectra of the different values of $n$: as $z$ increases, the angular power spectra for integral values of $n$ appear to group together (with relatively larger amplitude), and those for decimal values of $n$ appeared to group together (separately, with a relatively lower amplitude). This feature could be hunted down in the experimental data; if established, could serve as a ``smoking gun" for this kind or broken power-law $f(R)$ gravity models. (Future observational surveys will be able to detect such deviations in the large-scale structures.) In essence, within the context of modified gravity theories, it seems that relativistic effects will enhance the potential of using cosmic magnification as a cosmological probe.

We also compared the (standard) lensing magnification angular power spectrum in $f(R)$ gravity to that in $\Lambda$CDM. The results showed that lensing magnification in $\Lambda$CDM is stronger (or more enhanced) than in $f(R)$ gravity on all scales, at the given $z$. Moreover, on the largest scales, the magnitude of lensing magnification in $\Lambda$CDM falls relatively quickly (with increasing $z$) towards that of $f(R)$ gravity. Similarly, we compared the total (relativistic) magnification angular power spectrum in $f(R)$ gravity to that in $\Lambda$CDM. On the one hand, we found that, unlike in the lensing magnification scenario, here the relativistic corrections in $f(R)$ gravity will combine to give a significant net negative effect in the cosmic magnification at $z_S \,{\leq}\, 1$, which will substantially diminish the large-scale magnitude of the cosmic magnification, relative to $\Lambda$CDM (at the given $z$). On the other hand, at $z \,{>}\, 1$, the relativistic corrections will combine to give a net positive effect, thereby boosting the observed cosmic magnification in $f(R)$, relative to the $\Lambda$CDM.


\section*{Acknowledgements}
We thank the Centre for High Performance Computing, Cape Town, South Africa, for providing the computing facilities with which all the numerical computations in this work were done. AA acknowledges that this work is based on the research supported in part by the NRF with grant no.112131. AdlCD acknowledges support from NRF grants no.120390, reference: BSFP190416431035; no.120396, reference: CSRP190405427545; PID2019-108655GBI00, COOPB204064, I-COOP+2019 and PID2021-122938NB-I00 MICINN Spain. PKSD thanks First Rand Bank (South Africa) for financial support.

\section*{Data Availability}
Data sharing is not applicable to this article, as no datasets were generated or analysed in the current study. 




\bibliographystyle{mnras}
\bibliography{imprint_of_fofR_gravity-mnras} 


\appendix 

\section{$\lowercase{f}(R)$ Gravity: The Perturbations}\label{App:f(R)}
The gravitational field equations in this work are drawn from the comprehensive work by \cite{Clifton:2011jh}. 

For a universe composed of several fluid species $A$, the gravitational field constraint equations are given by 
\begin{align}\label{dpsi}
\Psi' + {\cal H}\Phi =&\; -\dfrac{1}{2}\kappa^2 a^2\sum_A \left(\bar{\rho}_A + \bar{p}_A\right) V_A,\\ \label{gradpsi}
\nabla^{2} \Psi - 3\mathcal{H} \left( {\cal H}\Phi +{\Psi}'\right) =&\; \dfrac{1}{2}\kappa^2 a^2 \sum_A{ \delta\rho_A},
\end{align}
where $\kappa \equiv 8{\pi}G$, with $V_x$ being given by
\bea\label{rhoxPxVx}
\kappa^2(\bar{\rho}_x+\bar{p}_x) V_x &{\equiv}& a^{-2}\left(\Phi f'_R + {\cal H}\delta{f}_R - \delta{f}'_R\right) \nn
&& +\; 2a^{-2}\left(\Psi' + {\cal H}\Phi\right)f_R , 
\eea
with $\bar{\rho}_x$ and $\bar{p}_x$ being as given in Sec.~\ref{subsec:backgrnd}, and the DE density perturbation is given by
\bea\label{deltaRho_x}
\kappa^2\delta\rho_x &{\equiv}& 2a^{-2}\left[3{\cal H}(\Psi'+{\cal H}\Phi) - \nabla^2\Psi\right]f_R \\ \nonumber
&& +\; a^{-2}\left(\nabla^2 + 3{\cal H}'\right)\delta{f}_R - 3a^{-2}{\cal H}\delta{f}'_R \nn
&& +\; 3a^{-2}\left(\Psi'+2{\cal H}\Phi\right)f'_R,
\eea
which give the velocity potential $V_x$ and density perturbation $\delta\rho_x$, respectively.

By combining \eqref{gradpsi} and \eqref{dpsi}, we have the Poisson equation, given by
\bea\label{PoissA}
\nabla^2\Psi \;=\; \dfrac{1}{2}\kappa^2 a^2 \sum_A \bar{\rho}_A\Delta_A ,
\eea
with the DE comoving overdensity $\Delta_x$, being given by
\bea\label{D_x}
\kappa^2 a^2 \bar{\rho}_x\Delta_x &=& 3\left(\Psi'+{\cal H}\Phi\right)f'_R - 2f_R\nabla^2\Psi \nn
&& +\; \left[\nabla^2+3\left({\cal H}'-{\cal H}^2\right)\right]\delta{f}_R,
\eea
where we used \eqref{Delta_A}, \eqref{rhoxPxVx} and \eqref{deltaRho_x}.

The Bardeen potentials \cite{Bardeen:1980kt} are related by
\bea\label{phipsiA} 
\Psi -\Phi \;=\; \kappa^2 a^2 \sum_A \left(\bar{\rho}_A + \bar{p}_A\right) \Pi_A ,
\eea
where we have
\bea\label{Pix}
\kappa^2\left(\bar{\rho}_x + \bar{p}_x\right) \Pi_x \;\equiv\; a^{-2}\left[\delta{f}_R + (\Phi - \Psi)f_R\right] ,
\eea
with $\bar{\rho}_x$ and $\bar{p}_x$ being as given in Sec.~\ref{subsec:backgrnd}.

Thus, given \eqref{dpsi} and \eqref{phipsiA}, we have the evolution of the temporal Bardeen potential to be determined by
\bea\label{dPhideta}
\Phi' &+& {\cal H}\Phi + \left[1+3c^2_{am}+\alpha_m + \dfrac{f'_R}{{\cal H}(1+f_R)}\right]{\cal H}\left(\Phi + \Psi\right) \nn
&=& -\; \dfrac{1}{2} \kappa^2 a^2 \sum_A \left(\bar{\rho}_A+\bar{p}_A\right)V_A - \dfrac{\delta{f}'_R}{1+f_R}\nn
&& -\; \left(1+3c^2_{am}+\alpha_m\right) \dfrac{{\cal H}\delta{f}_R}{1+f_R} ,
\eea
where $\alpha_m \equiv -\Pi'_m/({\cal H}\Pi_m)$.

The spatial Bardeen potential evolves according to the follwoing equation
\bea\label{psiA}
\Psi'' &+& 2{\cal H} \Psi' + {\cal H}\Phi' + \left({\cal H}^2 + 2{\cal H}'\right) \Phi \nn
&=& \dfrac{\kappa^2 a^2}{2} \sum_A \left[ \delta{p}_A +\dfrac{2}{3}\left(\bar{\rho}_A +\bar{p}_A\right) \nabla^{2}\Pi_A\right],
\eea
where 
\begin{align} \label{Pixdensity}
\kappa^2 \left[\delta{p}_x + \dfrac{2}{3}\left(\bar{\rho}_x+\bar{p}_x\right) \nabla^2\Pi_x\right]  {\equiv}&\; a^{-2}\Big\{ \delta{f}''_R + {\cal H}\delta{f}'_R - 2\Phi f''_R \nn
-& \left(4\nabla^2 +{\cal H}^2 + a^2R\right)\dfrac{\delta{f}_R}{6}  \nn
-&\; 2f_R\Big[\Psi'' + 2{\cal H} \Psi' + {\cal H}\Phi' \nn
&\hspace{1cm} +\; \left({\cal H}^2 + 2{\cal H}'\right) \Phi \nn
&\hspace{1cm} +\; \dfrac{1}{3}\nabla^2 (\Phi-\Psi)\Big] \nn
-& \left[\Phi' + 2(\Psi' + {\cal H}\Phi)\right]f_R' \Big\} ,
\end{align}
and we may use \eqref{Pix} to eliminate the $\Pi_x$ term---hence obtaining the DE pressure perturbation $\delta{p}_x$, as given by \eqref{delP_x}. By using \eqref{phipsiA} in \eqref{psiA}, we have
\begin{align}\label{psiA2}
\Psi'' + {\cal H} \left(2 + 3c^2_{a,\rm eff}\right)\Psi' +&\; {\cal H}\Phi' + \left[2{\cal H}' + \left(1+3c^2_{a,\rm eff}\right){\cal H}^2\right] \Phi \nn
=&\; \dfrac{1}{3}\nabla^2\left(\Psi - \Phi\right) + \dfrac{3}{2} {\cal H}^2 c^2_{s,\rm eff} \Delta_{\rm eff},
\end{align}
where we used the general expression for the pressure perturbation, for an arbitrary fluid, given by
\bea\label{delPx}
\delta{p}_A = c^2_{aA}\delta\rho_A + \left(c^2_{sA}-c^2_{aA}\right)\bar{\rho}_A\Delta_A ,
\eea
with $\Delta_A$ being given by \eqref{Delta_A}, and $c_{sA}$ is the physical sound speed of the species $A$---defined with respect to the rest frame of $A$; the effective adiabatic sound speed $c_{a,\rm eff}$ of the system:
\bea\label{c2aEff}
c^2_{a,\rm eff} &=& \dfrac{1}{1+w_{\rm eff}} \sum_A \Omega_A\left(1+w_A\right)c^2_{aA}, \nn
w_{\rm eff} &=& \sum_A \Omega_A w_A, 
\eea
with $w_{\rm eff} = -\left(1+2{\cal H}'/{\cal H}^2\right)/3 \,{=}\, (1-2x_3)/3$, being the effective equation of state parameter for the system, and $\Omega_A \,{\equiv}\, \bar{\rho}_A/\bar{\rho}_{\rm eff}$ are the energy density parameters---which evolve by
\bea\nonumber
\Omega'_A + 3{\cal H} \left(w_A-w_{\rm eff}\right)\Omega_A = 0,
\eea
where this includes DE, with $w_{\rm eff}$ as given by \eqref{c2aEff}. The effective physical sound speed $c_{s,\rm eff}$, which appears in \eqref{psiA2}, is given by
\bea\label{c2sEff}
c^2_{s,\rm eff} = \dfrac{1}{\Delta_{\rm eff}} \sum_A \Omega_A \Delta_A c^2_{sA},
\eea
where $\Delta_{\rm eff} = \sum_A \Omega_A \Delta_A$ is the effective (or total) comoving overdensity.

\section{Initial Conditions}\label{App:ICs}

\subsection{Background}
We set our initial conditions in the matter dominated epoch $z\,{=}\,z_i$. As stated in Sec.~\ref{subsec:model}, we use pseudo-$\Lambda$CDM initial conditions for the background: $w_{xi} \,{=}\, w_x(z_i) \,{=}\, {-}0.999$, and 
\bea\label{H_i}
h(z_i) = \Omega_{m0} a_i^{-3} + \left(1-\Omega_{m0}\right) a_i^{-3(1+w_{xi})}
\eea
where $h \,{\equiv}\, H/H_0$ and $a_i \,{=}\, a(z_i)$. We set $z_i = 20$, $x_3(z_i) \simeq 0.5005$ (inferred from $\Lambda$CDM at the same $z$); we used the definitions in \eqref{x_i} for $\bar{R}(z_i)$ [given the values of $h_i$ and $x_3(z_i)$ above], $x_2(z_i)$, and $x_4(z_i)$, accordingly; with $x_1(z_i)$ being estimated using the definition of $\Omega_x$ in \eqref{Omega_A} and the values of $x_2(z_i)$, $x_3(z_i)$, $x_4(z_i)$ and $\Omega_x(z_i)$ [from \eqref{H_i}].

The reason for the choice of $z_i \,{=}\, 20$ is two-fold: (1) This guarantees that the early universe is consistent with $\Lambda$CDM, because the given $z$ corresponds to the plateau of the Hu-Sawicki function \eqref{HuSawicki}---i.e., an effective cosmological constant, and (2) The background is devoid of any sudden curvature singularity issues, first pointed out by \cite{Frolov:2008uf} \citep[see also][]{Clifton:2015ira}.

Given the numerical complexity in handling the full system of equations of $f(R)$ theories, and also in order for us to fix the background to a single evolution (see discussion in Sec.~\ref{subsec:imprint}, first paragraph), we solved the background system separately (from $z_i \,{=}\, 20$ until today, $z \,{=}\, 0$, with the model exponent $n \,{=}\, 1$) and fit the solutions to analytical expressions. \citep[See also][where the phase space analysis for the Hu-Sawicki model was performed with $n \,{=}\, 1$.]{MacDevette:2022hts}

\subsection{Perturbations}\label{subsec:pertICs}
Similarly, as stated in Sec.~\ref{subsec:model}, we use adiabatic initial conditions for the perturbations. We adopt the Einstein de Sitter initial condition,
\bea\label{EdeS} 
\Psi'(z_i) = 0,
\eea
given that $\Omega_x(z_i)\ll 1$ for $20 \leq z_i \leq z_d$, and we assume that
\begin{align*}
&& \Omega_x(z_i) =&\; \Omega_x(z_d), & \Psi'(z_i) =&\; \Psi'(z_d),\nn
&& \Psi(z_i) =&\; \Psi(z_d), & \Phi(z_i) =&\; \Phi(z_d),
\end{align*}
where $z_d$ is the redshift at the photon-matter decoupling, with $\Psi(z_d) \,{=}\, \Psi_d$ as given by \eqref{Phi_d}.

We adopt adiabatic initial conditions, which are usually imposed by the vanishing of the relative entropy perturbation $S_{xm}$ \citep{Duniya:2019mpr, Duniya:2015ths, Duniya:2013eta, Duniya:2015nva, Duniya:2015dpa}, given by $S_{mx}(z_i) \,{=}\, 0$; consequently, $\delta{\rho}_x(z_i)/\bar{\rho}'_x(z_i) \,{=}\, \delta{\rho}_m(z_i)/\bar{\rho}'_m(z_i)$. Then by choosing the velocities to be equal, given by
\bea\label{vels_d}
V_x(z_i) \;=\; V_m(z_i),
\eea
and by using \eqref{Delta_A}, we obtain
\bea\label{Dx_Dm_i}
\dfrac{\Delta_x(z_i)}{1 +w_x(z_i)} \;=\; \dfrac{\Delta_m(z_i)}{1 +w_m(z_i)},
\eea
where given \eqref{PoissA} and that $\Omega_x(z_i)\ll 1$, we have
\bea\label{Delta_mi}
\Delta_m(k,z_i) &\;=\;& -\dfrac{2y^2}{3\Omega_m} \Psi_d(k), \\
\label{V_mi:V_xi}
V_m(k,z_i) &\;=\;& \dfrac{-2}{3{\cal H} \left(1 + w_{\rm eff}\right)} \Phi_d(k),
\eea
where $y \equiv k/{\cal H}$. The initial values $\delta{f}'_R(k,z_i)$, $\delta{f}_R(k,z_i)$ and $\Phi_d(k)$ may be obtained from \eqref{V_x}, \eqref{Delta_x} and \eqref{Phi_Psi}, respectively.


\bsp	
\label{lastpage}
\end{document}